\documentclass[12pt]{article}
\usepackage{amstext}
\usepackage{bm}
\usepackage{amssymb,amsmath,mathrsfs,setspace}
\usepackage{graphicx,psfrag,epsf}
\usepackage{array}
\usepackage{natbib}
\usepackage{epsfig}
\usepackage{amssymb}
\usepackage{url}
\usepackage{verbatim}
\usepackage{comment}
\usepackage{color}
\usepackage{graphicx,psfrag,subfigure,amsmath,amssymb,appendix,multirow,float,threeparttable,abstract}
\newtheorem{theorem}{Theorem}[section]
\newtheorem{lemma}[theorem]{Lemma}

\usepackage[top=1in, bottom=1in, left=1in, right=1in]{geometry}
\fontsize{12}{0}
\numberwithin{equation}{section}
\date{}

\begin{document}


\newcommand{\blind}{0}



\if0\blind
{
	\title{\bf eQTL Mapping via Effective SNP \\
		Ranking and Screening \thanks{Address for correspondence: X. Jessie Jeng, Department of Statistics, North Carolina State University, SAS Hall, 2311 Stinson Dr., Raleigh, NC 27695-8203, USA. E-mail: xjjeng@ncsu.edu. 
		}\hspace{.2cm}}
	\author{Jacob Rhyne$^1$, Jung-Ying Tzeng$^{1,2,3,4}$, Teng Zhang$^1$, and X. Jessie Jeng$^1$ \\ 
		1. Department of Statistics, North Carolina State University\\
		2. Bioinformatics Research Center, North Carolina State University \\
		3. Department of Statistics, National Cheng-Kung University, Tainan, Taiwan. \\
		4. Institute of Epidemiology and Preventive Medicine,\\ 
		National Taiwan University, Taipei, Taiwan\\
		}
	\maketitle
} \fi

\if1\blind
{
	\bigskip
	\bigskip
	\bigskip
	\begin{center}
		{\LARGE\bf eQTL Mapping via Effective SNP Ranking and Screening \\}
	\end{center}
	\medskip
} \fi

\hspace{-0.5in}


\begin{abstract}
Genome-wide eQTL mapping explores the relationship between gene expression values and DNA variants to understand genetic causes of human disease. Due to the large number of genes and DNA variants that need to be assessed simultaneously, current methods for eQTL mapping often suffer from low detection power, especially for identifying \textit{trans}-eQTLs. In this paper, we propose a new method that utilizes advanced techniques in large-scale signal detection to pursue the structure of eQTL data and improve the power for eQTL mapping. The new method greatly reduces the burden of joint modeling by developing a new ranking and screening strategy based on the higher criticism statistic. Numerical results in simulation studies demonstrate the superior performance of our method in detecting true eQTLs with reduced computational expense.  The proposed method is also evaluated in HapMap eQTL data analysis and the results are compared to a database of known eQTLs.
\end{abstract}


\section{Introduction} \label{s:intro}

Expression quantitative trait loci (eQTLs) are genomic regions that carry DNA sequence variants that influence gene expression \citep{review2015}. eQTLs act either in \textit{cis}, if the SNP is located near the gene whose expression is influenced, or in \textit{trans}, if the SNP is located further away from the gene \citep{SNP_cis_trans, eQTL_mapping_cookson}.  \textit{trans}-eQTLs are particularly challenging to identify in human population studies because of their weak effects \citep{westra2013}.

\begin{figure} [h]
	\label{fig:eqtl_example}
	\centering
	\caption{An example of eQTLs from \cite{wolen2011identifying}}
	\includegraphics[scale = 0.4]{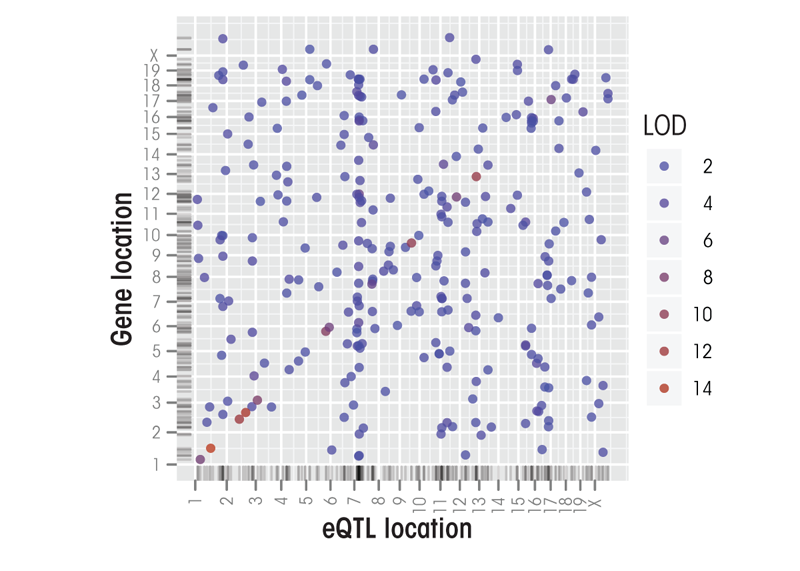}
\end{figure}

Genome-wide eQTL mapping, first proposed by Jansen and Nap in 2001, explores the relationship between gene expression levels and DNA variants, including single nucleotide polymorphisms (SNPs), copy number variants (CNVs), and short tandem repeats and single amino acid repeats \citep{eQTL_mapping_cookson}.  The goal of eQTL mapping is to identify genetic variants, usually SNPs, that are significantly associated with the expression of genes \citep{shabalin2012matrix}.  Figure \ref{fig:eqtl_example} shows an example of eQTL results from \cite{wolen2011identifying}. In this figure, each dot denotes an identified significant association between a gene and a SNP. The results reveal that the expression of a gene can be associated with multiple SNPs, suggesting there exist potential joint effects of multiple SNPs on the expression value of a gene. In addition, some SNPs are strongly associated with a few genes, (e.g., red dots), and in this paper, we refer to them as strong-sparse eQTLs. Some are moderately associated with a number of genes, (e.g., a SNP with multiple blue dots), and we refer to them as hotspot eQTLs or weak-dense eQTLs.  

eQTL mapping plays a pivotal role in human disease research, and can help to elucidate underlying genetic mechanisms of diseases \citep{eQTL_mapping_cookson, eQTL_Present_Future}.  Many links between genetic markers, such as SNPs, and diseases have been found through genome-wide association studies (GWAS). eQTL mapping provides a link to understand how SNPs actually influence diseases; when SNPs are associated with the expression of a gene in eQTL mapping and with a disease in GWAS, this may imply that the expression of the gene mediates the SNP effect on the disease \citep{review2015}. eQTL mapping has helped to elucidate the connection between SNPs and diseases such as Type 1 diabetes, ulcerative colitis, Chron's disease, and many autoimmune diseases, especially when SNPs are in non-coding regions \citep{review2015}.

In traditional eQTL analysis, the SNP effect on gene expression is assessed marginally, i.e. one SNP-gene pair at a time, which leads to a large number of tests \citep{shabalin2012matrix}. The significant gene-SNP pairs are identified by controlling type I error rates, for example, the family-wise error rate or the false discovery rate \citep{benjamini1995controlling}. This marginal strategy has the advantage of being computationally efficient and easy to apply. However, marginal eQTL selection has two primary drawbacks: (1) The total number of marginal tests, i.e., the number of SNPs times the number of genes, is large and multiplicity adjustment tends to result in a conservative finding, especially for identifying \textit{trans}-eQTLs and eQTL hotspots. (2) Marginal selection overlooks the potential joint effect of SNPs on gene expression and cannot account for the non-genetic effects. Ignoring these two features may not precisely capture the underlying SNP-expression relationship and lead to a loss of power. These flaws of marginal eQTL selection have motivated the consideration of joint modeling methods that explore the relationship of gene expression and SNPs simultaneously.


There are many joint modeling methods available for eQTL analysis \citep{PANAMA, StatMethods, LMMEH, yang2013LORS}.  However, without first performing screening to reduce the number of SNPs, joint modeling is often computationally infeasible \citep{yang2013LORS}.  For example, the authors of LMM-EH \citep{LMMEH}, an earlier method performing joint modeling while accounting for hidden factors, reported that an analysis of a mice dataset required 10 hours to complete even when parallelized across 1,100 processors.  LORS \citep{yang2013LORS}, another popular joint modeling method that accounts for hidden factors, requires $O(qn^2 + n^2pq)$ operations to run, where $n$ is the sample size, $q$ is the number of genes, and $p$ is the number of SNPs.  Since human datasets often have $O(10^4)$ genes and $O(10^6)$ SNPs, LORS is computationally very expensive for genome-wide eQTL mapping in human datasets.  In order to make joint modeling feasible for large datasets, screening methods can be used to reduce the number of SNPs.

Since only a fraction of SNPs influence gene expression, an optimal screening method will remove the unimportant SNPs, i.e., SNPs not associated with expression, while keeping the important SNPs.  The results of the screening greatly impact the joint modeling.  If the screening method is not powerful enough, weak eQTLs may be lost too early and not have a chance to be detected with joint modeling.  On the other hand, if the number of SNPs removed by screening is too conservative, there may be too many SNPs for joint modeling to be computationally feasible.  There is a great need for a screening method that is powerful enough to keep weak eQTLs but still reduce the number of SNPs enough so that the overall computation complexity is reasonable.


A classic example of screening is to use linear regression to model the marginal relationship between each SNP-gene pair, 
record the p-value of the regression coefficient, and select those SNPs that have at least one p-values less than a pre-specified threshold \citep{marginal_SNP_selection}.  The SNPs selected are used in further analysis such as joint modeling.  This classical method and other similar screening methods that select those SNPs associated with at least one genes have three main flaws:  (1) The threshold for marginal $p$-values is often chosen arbitrarily or by convention; (2) these methods generally struggle to identify weak effects of \textit{trans}-eQTLs; and 
(3) the existing screening criteria cannot directly indicate the computation complexity of the follow-up joint modeling.  
Additionally, these methods have a tendency to carry over a large number of SNPs that do not actually influence gene expression into joint modeling, which affects the efficiency of joint modeling. 

In this paper, we propose a new ranking and screening strategy to largely reduce the number of SNPs before joint modeling while retaining both strong and weak eQTLs.  Instead of ranking SNP-gene pairs, we propose to rank only SNPs utilizing the Higher Criticism (HC) statistic  that summarizes the association of the SNP to all genes. The HC statistic has been developed in high-dimensional statistics for simultaneous detection of sparse and dense signals \citep{donoho2004higher, HC2011}. In eQTL mapping, an important SNP may be strongly associated with a few genes (strong-sparse) or weakly associated with a number of genes (weak-dense). Ranking the SNPs by their HC statistics favors both strong-sparse and weak-dense SNPs over noise SNPs. 

Consequently, a screening procedure can be built upon the HC ranking to efficiently retain strong-sparse and weak-dense SNPs for joint modeling.
We suggest to restrict the number of SNPs that are allowed to be carried over to joint modeling to the sample size $n$, so that joint modeling can be performed effectively with adequate sample size. Comparing to the existing screening methods that select the top-ranked SNP-gene pairs,  our method directly controls the total number of SNPs carried over  to joint modeling while effectively retains strong-sparse and weak-dense eQTLs, and greatly reduces the computation complexity.

After HC ranking and screening, many joint modeling methods can be applied to gene expression data and the reduced subset of SNPs.  We outline our methodology in detail in Section 2 and provide an overview of a candidate joint modeling method that can be used after screening.  Following this, we perform a simulation study to demonstrate the finite-sample performance of our method, and apply our method to eQTL data from the International HapMap Project.

\section{Methodology}

\subsection{Initial Estimate}

The first step of our method is to construct an initial estimate $\hat{\textbf{B}}_{p\times q}$, where $p$ and $q$ are the numbers of SNPs and genes, respectively. The initial estimate can be obtained by the marginal regression method introduced in \citet{yang2013LORS} that solves the optimization problem   

\begin{equation} 
	\label{LORS_Screening_optimization}
	min_{(\beta_i, \mu, \textbf{L})} \frac{1}{2} \lVert \textbf{Y} - \textbf{X}_i\beta_i - \textbf{1}\mu - \textbf{L} \rVert_F^2 + \lambda \lVert \textbf{L} \rVert_*,
\end{equation}
where \noindent $\textbf{Y} \in \mathbb{R}^{n \times q}$ is a matrix of gene expression levels, $\textbf{X}_i \in \mathbb{R}^{n \times 1}$ is the vector of genotypes of the $i^{th}$ SNP, $\beta_i \in \mathbb{R}^{1 \times q}$ is a row vector of coefficients corresponding to the association of the $i^{th}$ SNP to all the $q$ genes, $\mu \in \mathbb{R}^{1 \times q}$ is a row vector of intercepts, 
$\textbf{L} \in \mathbb{R}^{n \times q}$ is a matrix of non-genetic factors that influence gene expression, and $\lVert \textbf{L} \rVert_*$ is the nuclear norm of {\bf L}.  
The optimization problem in equation (\ref{LORS_Screening_optimization}) provides a mechanism for handling hidden, non-genetic factors that influence gene expression.  Accounting for these hidden, unobservable factors has been shown to lead to better eQTL mapping performance \citep{PANAMA,LMMEH,yang2013LORS}.  The algorithm to solve (\ref{LORS_Screening_optimization}) is presented in the appendix.  Denote the estimate for the coefficients of the $i^{th}$ SNP as $\hat{\beta}_i \in \mathbb{R}^{1 \times q}$ for $i=1,\dots,p$.  We stack these $p$ row vectors to form the initial estimate $\hat{\textbf{B}}_{p\times q}$.

\subsection{Summary Statistic for each SNP using Higher Criticism}

The next phase of our methodology is to rank each SNP by its importance to the gene expression levels.  We propose to use the higher criticism (HC) statistic to measure the importance of each SNP because the HC statistic has a desirable property of being sensitive to both sparse and dense signals \citep{HC2011}.

In order to construct the HC statistic, the initial estimate $\hat{\textbf{B}}_{p \times q}$ must be standardized.  Let $Z_{ij}$ be the standardized $\hat{\beta}_{ij}$ corresponding to the $i^{th}$ SNP and the $j^{th}$ gene. We calculate the standardized estimates $Z_{ij}$ as
\begin{equation}
	\label{std_est}
	Z_{ij} = \frac{\hat{\beta}_{ij}}{\sqrt{(\textbf{X}_i^T\textbf{X}_i)^{-1}\hat{V}(\textbf{Y}_j - \textbf{X}_i\hat{\beta}_{ij})}},
\end{equation} 
where $\hat{V}(\textbf{Y}_j - \textbf{X}_i\hat{\beta}_{ij})$ denotes the sample variance of $\textbf{Y}_j - \textbf{X}_i\hat{\beta}_{ij}$.

For a given SNP, the null hypothesis is that the SNP is not associated with any genes.  The alternative hypothesis is that the SNP is associated with at least one gene.  The hypotheses for all SNPs can be formulated as follows for $i=1,\dots,p$,

\begin{equation}
	\label{Hyp_Test}
	\begin{split}
		H_{0i}: Z_{ij} \sim N(0, 1), \hspace{4.3cm} j = 1, \ldots, q,  
		\\
		H_{1i} : Z_{ij} \sim (1 - \eta_i) N(0, 1) + \eta_i N(\mu_{i}, \sigma_{i}^2), \quad j = 1, \ldots, q,
	\end{split}
\end{equation}

\noindent where $\eta_i \in [0,1]$ denotes the proportion of genes regulated by SNP $i$; $\mu_{i}$ $(\ne 0)$ and $\sigma_{i}^2$ $(> 0)$ are the mean and variance of $Z_{ij}$, respectively, if SNP $i$ is associated with gene $j$.  All of the parameters $\eta_i$, $\mu_{i}$, and $\sigma_{i}$ are unknown and varying with $i$.

In reality, most SNPs are not truly associated with any genes; therefore most of the null hypotheses $H_{0i}, i = 1, \ldots, p$, are true. 
The alternative hypotheses describe the distribution of $Z_{ij}$ for important SNPs by mixture models.  Such mixture models include the strong-sparse case ($\mu_i$ is relatively large and $\eta_i$ is small) and the weak-dense case ($\mu_i$ is small and $\eta_i$ is relatively large).

Donoho and Jin (2004) developed the HC statistic to test the existence of sparse signals among a large number of observations. In this paper, we propose to use the HC statistic to summarize the importance of each SNP in regulating the genes. We present an analogous version of the HC statistic based on the standardized test statistic $Z_{ij}$.  Let $S_i(t) = \sum_{j = 1}^q I\{|Z_{ij}|\geq t\}$.  The HC statistic of SNP $i$ is calculated as 

\begin{equation}
	\label{hc}
	HC_i = \sup_{t \geq 0} \bigg\{ \sqrt{q} \frac{S_i(t)/q - \bar{\Phi}(t)}{\sqrt{\bar{\Phi}(t)(1 - \bar{\Phi}(t))}} \bigg\}. \nonumber
\end{equation}

Note that under the null hypotheses in equation (\ref{Hyp_Test}), the expectation and variance of $S_i(t)/q$ are 

\begin{equation}
	\label{exp_var}
	E(S_i(t)/q) = \bar{\Phi}(t); \quad V(S_i(t)/q) = \frac{\bar{\Phi}(t)(1 - \bar{\Phi}(t))}{q}, \nonumber
\end{equation}

\noindent where $\bar{\Phi}(t) = P(N(0, 1) > t)$. Therefore large values of $HC_i$ suggest that SNP $i$ is important in regulating the genes. More specifically, when SNP $i$ is important ($H_{1i}$ is true), $Z_{ij}$ has non-zero mean value for some $j$, which causes elevated mean of $S_i(t)$ for some $t$. Consider the case that SNP $i$ is strong-sparse, we observe several large $Z_{ij}$, and the standardized $S_i(t)$ may reach its max at a large $t$. On the other hand, when SNP $i$ is weak-dense, the standardized $S_i(t)$ may reach its max at a moderate $t$. By taking the maximum over $t$, the HC statistic can capture the information for both strong-sparse and weak-dense SNPs.

\subsection{HC Ranking and Screening}

After the HC statistics are calculated for all SNPs, we propose to rank the SNPs by their HC statistics in a decreasing order.
Since ranking SNPs by their HC statistics tends to place strong-sparse and weak-dense SNPs before irrelevant SNPs, we can select only the top-ranked SNPs for joint modeling. We may select the top $n$ ranked SNPs,  where $n$ is the sample size, so that the ratio of sample size to dimensionality is adequate for joint modeling. Similar screening criterion has been used in 
\cite{yang2013LORS}. However, \cite{yang2013LORS} does not rank SNPs by a summary statistic; they rank and keep the top $n$ SNPs by the initial estimate $\hat{\textbf{B}}$ for each gene. The SNPs entering joint modeling are the union of the SNPs selected for each gene. Compared to \cite{yang2013LORS}, our ranking and screening strategy can be more efficient in selecting much less SNPs for joint modeling while retaining both strong-sparse and weak-dense SNPs.

\subsection{Joint Modeling after SNP Screening}

Let $\textbf{X}_r \in \mathbb{R}^{n \times n}$ represent the reduced SNP matrix corresponding to the $n$ SNPs selected by the HC ranking and screening.  A joint modeling method can be used to explore the relationship between the gene expression matrix $\textbf{Y}$ and $\textbf{X}_r$.   A popular method is LOw-Rank representation to account for confounding factors and make use of Sparse regression for eQTL mapping (LORS) \citep{yang2013LORS}, the model structure for which is

\begin{equation}
	\label{LORS_Model}
	\textbf{Y}= \textbf{1}\textbf{$\mu$} + \textbf{X}_r\textbf{B}_r + \textbf{L} + \textbf{e}, \nonumber
\end{equation}
where $\textbf{Y} \in \mathbb{R}^{n \times q}$ is the matrix of gene expression levels, $\textbf{X}_r \in \mathbb{R}^{n \times n}$ is the reduced matrix of SNPs, $\textbf{B} \in \mathbb{R}^{n \times q}$ is a matrix of coefficients, $\textbf{L} \in \mathbb{R}^{n \times q}$ is a matrix of non-genetic factors, $\textbf{1} \in \mathbb{R}^{n \times 1}$ is a vector of ones, $\textbf{$\mu$} \in \mathbb{R}^{1 \times q}$ is a row vector of intercepts, and $\textbf{e}\in \mathbb{R}^{n \times q}$ is an error matrix with $e_{ij} \sim N(0,\sigma^2)$.  It is assumed that there are only a few hidden, non-genetic factors that influence gene expression and that there are only a small fraction of true SNP-gene associations. In other words, $\textbf{L}$ is low-rank and $\textbf{B}_r$ is sparse.  

In Lagrangian form, the LORS optimization problem is

\begin{equation}
	\label{LORS_Opt}
	min_{\textbf{B}, \mu, \textbf{L}} || \textbf{Y} - \textbf{X}_r\textbf{B}_r - \textbf{1}\textbf{$\mu$} - \textbf{L}||^2_F + \rho||\textbf{B}_r||_1 + \lambda||\textbf{L}||_*,
\end{equation}
where $\rho$ enforces the sparsity constraint on $\textbf{B}_r$ and $\lambda$ enforces the low-rank constraint on $\textbf{L}$.  The LORS MATLAB implementation is freely available at \url{http://zhaocenter.org/software/}. We also implemented LORS algorithm in R, which is available upon requests from the authors.

Besides LORS, there are other methods that can be applied at the joint modeling stage,  giving the users freedom to select the method that best suits their needs; for example \cite{LMMEH, PANAMA, VR2013, HEFT, Confetti}; etc.

\section{Simulation Study}
\label{s:sim_study}

In this section, we perform simulation studies to demonstrate the efficiency of HC ranking and compare HC ranking with other popular ranking methods.  The SNPs selected by HC ranking and screening are carried over to joint modeling. We compare the whole procedure with the marginal screening and joint modeling procedure in \citet{yang2013LORS}.

\subsection{Simulation Setup}

Genotype data were obtained from HapMap3, the third phase of the International HapMap Project. We focus on the data of chromosome 1, which includes  $24806$ SNPs for 160 subjects after LD pruning. Details of the data are provided in Section \ref{sec:realdata} (eQTL Data Analysis). We simulate expression data of 200 genes for the 160 subjects.

The setup of the simulation study is similar to the design of the simulation study in the original LORS paper \citep{yang2013LORS}.  Let $\textbf{B} \in \mathbb{R}^{24806 \times 200}$ be the coefficient matrix of SNP effects and let $\textbf{U} \in \mathbb{R}^{160 \times 200}$ be a matrix of non-genetic factors.  Since we assume that $\textbf{B}$ is sparse, only 20 SNPs were set to be active, and for each active SNP, $m=10$ or $m=50$ genes were randomly selected to be influenced.  We set $\beta$ to be either 0.5 or 2.  The case where $\beta=2$ and $m=10$ represents the strong-sparse scenario and the case where $\beta=0.5$ and $m=50$ represents the weak-dense scenario.  

It is known that hidden, non-genetic factors influence gene expression \citep{PANAMA, Confetti, LMMEH, yang2013LORS}.  We simulate a matrix of hidden factors \textbf{U} through the following steps: (1) $\textbf{H}_{n\times k} \sim N(0, I)$ where $k$ is the number of hidden factors and is set to 10; (2) $\textbf{$\Sigma$}_{n \times n} = \textbf{H}\textbf{H}^T$; (3) $\textbf{U}_j \sim N(\textbf{0},0.1 \times \textbf{$\Sigma$})$, where $\textbf{U}_j$ is the $j^{th}$ column of \textbf{U}.  Finally, $\textbf{Y}$ was simulated by

\begin{equation}
	\label{sim_setup}
	\textbf{Y} = \textbf{XB} + \textbf{U} + \textbf{e}, \nonumber
\end{equation}

\noindent where $\textbf{e}$ has its $j^{th}$ column simulated by $e_j \sim N(\textbf{0},\textbf{I})$.  

\subsection{Effectiveness of Ranking SNPs by HC Statistic} \label{sec:sim_ranking}

First, we evaluate the effectiveness of the HC ranking.  It is critical that an effective ranking procedure would rank active SNPs highly.  Two baseline methods are used in comparison to the HC ranking:  (1) ranking SNPs by the means of each row of $\hat{\textbf{B}}$ and (2) ranking SNPs by maximum absolute value by row.  We refer to these two alternative ranking procedures as ROWMEANS and EXTREMEVAL respectively.  By design, ROWMEANS should be good at detecting weak-dense SNPs and EXTREMEVAL should be good at detecting strong-sparse SNPs.

\begin{figure}[h]
	\centering
	\caption{Average precision versus recall for SNP ranking. The left plot illustrates the strong-sparse case with $\beta=2, m=10$. The right plot illustrates the weak-dense case with $\beta=0.5, m=50$. In the legend, HC represents the higher criticism ranking, RM the ROWMEANS ranking, and EV the EXTREMEVAL ranking.}
	\includegraphics[width=80mm, height = 75mm]{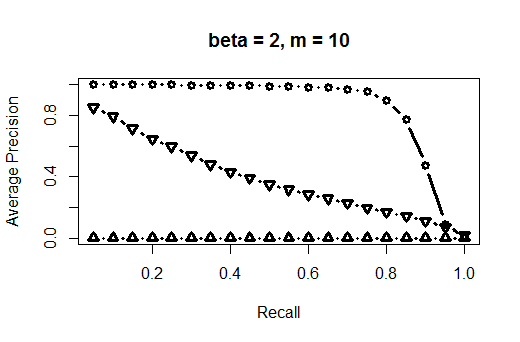}  
	\includegraphics[width=80mm, height = 75mm]{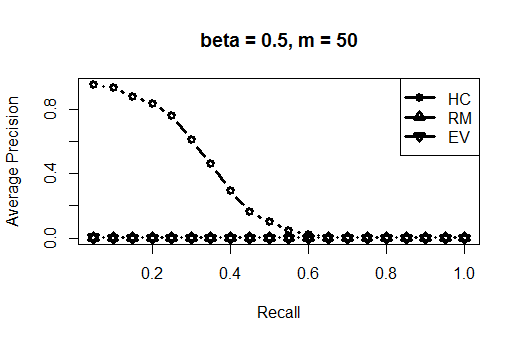} 
	\label{fig:sim_PR}
\end{figure}

To evaluate the effectiveness of the three ranking methods, we use precision-recall curves.  A precision-recall curve, commonly used in information retrieval, shows the precisions of a selection rule for different values of recall. The precision and recall are defined as
\begin{equation}
	\label{prec_rec}
	Precision=\frac{TP}{TP+FP} \qquad Recall=\frac{TP}{TP+FN}, \nonumber
\end{equation}
\noindent where $TP$ is the number of true positive SNPs, $FP$ is the number of false positive SNPs, and $FN$ is the number of false negative SNPs. 
Recall takes values in $\{1/20, 2/20, \ldots, 20/20\}$ because there are 20 active SNPs. For a given recall value, we calculate the average precisions (over 100 simulations) for all three ranking methods and report the results in Figure \ref{fig:sim_PR}. A method with higher precision-recall curve ranks more active SNPs before noise SNPs, relative to the other methods. 

As seen in Figure \ref{fig:sim_PR}, the HC ranking performs better at detecting and prioritizing active SNPs over noise SNPs than either the ROWMEANS ranking or EXTREMEVAL ranking in both strong-sparse and weak-dense settings.  The ROWMEANS method performs particularly poorly, with a very low precision for each value of recall.  The EXTREVEVAL ranking is more competitive with the HC ranking in the strong-sparse case, but the HC ranking still maintains an advantage.

\subsection{Screening and Joint Modeling}

In this section, we compare the performance of our procedure with the marginal screening and joint modeling procedure in  \cite{yang2013LORS}. For fairness of comparison, LORS algorithm is used for joint modeling in both procedures. We refer to the procedure in \cite{yang2013LORS} as MS-LORS and 
our multi-step procedure as HC-LORS that provides eQTL mapping by a) using the algorithm in equation (\ref{LORS_Screening_optimization}) to provide an initial estimate; b) standardizing the estimate and calculating HC statistics; c) ranking SNPs by HC and selecting the top $n$ SNPs; and d) applying LORS algorithm for joint modeling on the selected SNPs and the expression matrix.  
We compare both methods by their success in detecting true eQTLs and the computation time required to run each method.

\begin{figure} [h]
	\centering
	\caption{Average precision for the top 1,000 associations in strong-sparse (left) and weak-dense (right) scenarios.  In the legend, HC-LORS refers to the proposed multi-step procedure and MS-LORS refers to the screening and modeling method in \cite{yang2013LORS}.}
		\includegraphics[width=80mm, height=75mm]{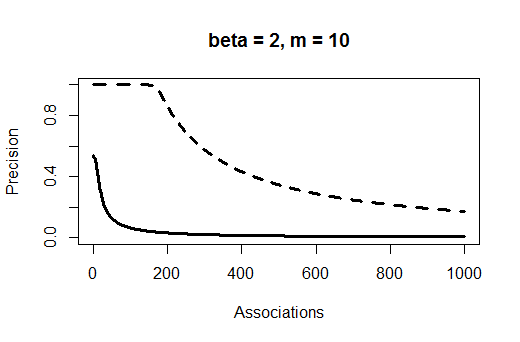} 
		\includegraphics[width=80mm, height=75mm]{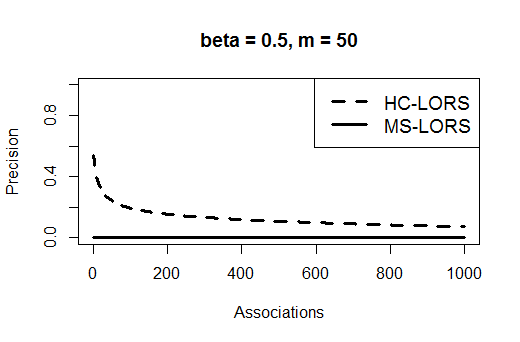}
		\label{fig:sim_precision}
\end{figure}

Denote the estimates of $\textbf{B}$ from HC-LORS and MS-LORS as $\hat{\textbf{B}}_{HC}$ and $\hat{\textbf{B}}_{MS}$, respectively.  The entries of $\hat{\textbf{B}}_{HC}$ and $\hat{\textbf{B}}_{MS}$ are ranked by magnitude in descending order.  For each $\hat{\textbf{B}}_{HC_{ij}}$ or $\hat{\textbf{B}}_{{MS}_{ij}}$ that is nonzero, if the corresponding true $\textbf{B}_{ij} \ne 0$ the association is called a true positive (TP); if $\textbf{B}_{ij} = 0$, the entry is called a false positive (FP).  The precision for the largest 1,000 associations sorted by the estimates of $\textbf{B}_{ij}$ is presented in Figure 3.  Higher precision means a larger proportion of true eQTLs are captured in the selected SNP-gene association pairs.
As seen in Figure \ref{fig:sim_precision}, HC-LORS performs better than MS-LORS in detecting true eQTLs and prioritizing them over noise in both strong-sparse and weak-dense scenarios.

HC-LORS requires less computational expense than MS-LORS due to effective ranking and screening.  In fact, the screening step of MS-LORS selects 3066 and 3059 SNPs on average in the strong-sparse and weak-dense scenarios, respectively, whereas our HC ranking and screening selects only $160$ SNPs in both scenarios.  
Consequently, the computation time of HC-LORS for joint modeling is much less than that of MS-LORS. Table \ref{tab:sim_time} summarizes the median values of the computation times for joint modeling of the two methods from 100 replications. 
The improvement in computational cost coupled with more accurate eQTL detection makes HC-LORS a promising method for large scale eQTL analysis.

\begin{table}
	\caption{Median values of computation time (in seconds) for joint modeling} \label{tab:sim_time}
	\begin{center}
		\begin{tabular}{ l  l  l }
			\hline
			Setting & Method & Computation time \\ \hline
			strong-sparse $(\beta=2, m=10)$ & HC-LORS & 509 \\  
			& MS-LORS & 1423 \\ \hline
			weak-dense $(\beta=0.5, m=50)$ & HC-LORS & 581 \\ 
			&  MS-LORS & 1223  \\ \hline
		\end{tabular}
	\end{center}
\end{table}

\section{eQTL Data Analysis} \label{sec:realdata}

We illustrate the utility of HC-LORS on the expression and SNP data from the third phase of International HapMap Project (HapMap3).  The dataset includes 1301 samples from a variety of human populations. The genotype data are provided on \url{ftp://ftp.ncbi.nlm.nih.gov/hapmap/genotypes/hapmap3_r3/plink_format/} and the details of gene expression data are provided on \url{http://www.ebi.ac.uk/arrayexpress/experiments/E-MTAB-264/}. 
We focus on chromosome 1 and use PLINK to remove correlated SNPs via linkage disequilibrium pruning with a window size 50, a moving window increment of 5 SNPs, and a cutoff value of $r^2 = 0.5$.  Again using PLINK, the pruned genotype data were converted into numerical SNP data.  Following this, the data were set to include only Asian individuals for which there were gene expression data available.  The resulting data have 2010 gene probes and 24806 SNPs for 160 subjects. Although the influence of tuning the parameters $\lambda$ and $\rho$ in equation (\ref{LORS_Opt}) should be minor, we address potential issues by using cross validation to select the parameters in (\ref{LORS_Opt})  \citep{monte_carlo_cross_valid,yang2013LORS}.

We use the classification method of Westra et al. (2013) to identify eQTLs as \textit{cis} or \textit{trans} \citep{westra2013}.  We call an eQTL \textit{cis} if the distance between the base pair positions of the SNP and the probe midpoint is less than 250 kilobases (kb) and call the eQTL \textit{trans} if this distance is greater than 5 megabases (mb).  
The top 5 SNPs, in terms of the estimated association effects, along with the associated probes are seen in Table \ref{tab:data_top}. 
A complete list of identified SNP and probe associations for the two methods, ordered by the estimated association effects, can be found in the supplemental document.

\begin{table}[ht]
	\caption{SNPs with the strongest estimated associations and the distance between the SNP location and probe midpoint.  HC-LORS denotes the proposed multi-step procedure; MS-LORS denotes the screening and modeling method in \cite{yang2013LORS}.} \label{tab:data_top}
	\begin{tabular}{p{1.75cm} llll}
		\hline
		\textbf{Method} & \textbf{SNP (Gene)} & \textbf{Probe (Gene)} & \textbf{Distance} & \textbf{Class.} \\ \hline
		\multirow{5}{*}{HC-LORS} & rs12745189 (Intergenic) & ILMN\_1762255 (GSTM1) & 45.34 (kb) & cis  \\
		& rs12745189 (Intergenic) & ILMN\_1668134 (GSTM1) & 46.55 (kb) & cis  \\
		& rs4657741 (TIPRL) & ILMN\_1779432 (TIPRL) & 21.53 (kb) & cis  \\
		& rs2138686 (BMP8B) & ILMN\_1653730 (OXCT2) & 7.68 (kb) & cis  \\
		& rs3007708 (Intergenic) & ILMN\_1796712 (S100A10) & 12.23 (kb) & cis  \\ \hline
		\multirow{5}{*}{MS-LORS} & rs2239892 (GSTM1) & ILMN\_1762255 (GSTM1) & 2.38 (kb) & cis  \\
		& rs10802457 (ZNF695) & ILMN\_1705078 (CHI3L2) & 133.63 (mb) & trans  \\
		& rs2239892 (GSTM1) & ILMN\_1668134 (GSTM1) & 1.18 (kb) & cis  \\
		& rs10802457 (ZNF695) & ILMN\_1685045 (CHI3L2) & 133.64 (mb) & trans  \\
		& rs12121466 (Intergenic) & ILMN\_1769839 (L1TD1) & 44.50 (mb) & trans  \\
		\hline
	\end{tabular}
\end{table}

As seen in Table \ref{tab:data_top}, the strongest associations detected by HC-LORS represent \textit{cis}-eQTLs, as the distance between the probe midpoint and the SNP position is small, and the strongest associations detected by MS-LORS are \textit{cis} and \textit{trans}-eQTLs.  Since \textit{trans}-eQTLs tend to have very weak signals, we expect the strongest signals to correspond to \textit{cis}-eQTLs \citep{westra2013}.  Both methods detect a large proportion of \textit{trans}-eQTLs; out of the 158 associations detected by HC-LORS, 113 of them are classified as potential \textit{trans}-eQTLs and out of the 117 associations detected by MS-LORS, 106 are classified as \textit{trans}-eQTLs.  This suggests that both methods show promise in \textit{trans}-eQTL detection.  

In addition to identifying candidate \textit{cis} and \textit{trans}-eQTLs, both methods identify SNPs associated with multiple genes, or hotspots.  Since Yang et al. (2013) define hotspot as a SNP being associated with the expression of at least 15 genes out of 7084 (or 0.21\%), we define hotspots as cases where a SNP is associated with at least $\lceil 0.0021 \times 2010 \rceil = 5$ genes. It can be seen in Table \ref{tab:data_hotspot} that HC-LORS detects four hotspots and MS-LORS detects two hotspots.  

\begin{table}[ht]
	\caption{SNPs that are declared hotspots by HC-LORS and MS-LORS} \label{tab:data_hotspot}
	\begin{tabular}{lll}
		\hline
		\textbf{Method} & \textbf{SNP (Gene) } & \textbf{Associated Genes}  \\ \hline
		HC-LORS & rs2297663 (LRP8) & AIM2 GSTM1 L1TD1 CAMK1G C1orf123 \\
		& & PHGDH  SH2D2A  \\ 
		& rs12408890 (WDR8) & LRRC8C WDR8 GSTM1 L1TD1 FAM79A \\
		& & GPX7  \\ 
		& rs12745189 (Intergenic) & GSTM1 GSTM1 GSTM4 GSTM4 HSPA6  \\ 
		& rs11204737 (ARNT) & CTSS ARNT CGN MNDA PRRX1  \\ \hline
		MS-LORS & rs4926440 (SCCPDH) & LRRC8C ATP1B1 IL23R SFN BCAR3 MFSD2 \\
		& & NECAP2 WDR8 RAP1GAP EDG1 PIGR \\
		& & IBRDC3 CNN3 GSTM1 GSTM1 RALGPS2 \\
		& & ZC3H12A L1TD1 MAP3K6 FLJ20054 IPP \\
		& & WNT3A C1orf63 CGN NPL ATP1B1 DARC \\
		& & PADI4 DENND2D FCGR2B VASH2 SCCPDH \\
		& & LOC645436 ATP2B4 LOC644094 C1orf54 \\
		& & SSR2 LOC391157 HSPA6 LOC729853 NMNAT2 \\
		& & FLJ25476 LPGAT1 TNFSF4  \\ 
		& rs2239892 (GSTM1) & LRRC8C WDR8 GSTM1 L1TD1 FAM79A GPX7  \\ \hline
	\end{tabular}
\end{table}

\begin{figure}
	\caption{Proportion of identified eQTLs that also appear in seeQTL database. The left plot presents the proportion for the ranked eQTLs of each method. The right plot zooms in on the top 25 detected eQTLs of each method. In the legend, HC-LORS and MS-LORS refer to our multi-step method and the screening and modeling method presented in \cite{yang2013LORS}, respectively. } \label{fig:data_verify}
	\begin{tabular}{cc}
		\includegraphics[width=80mm, height=75mm]{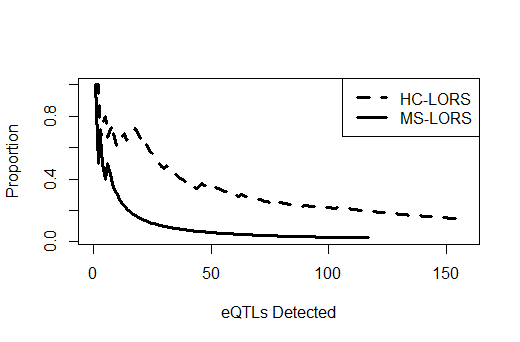} &
		\includegraphics[width=80mm, height=75mm]{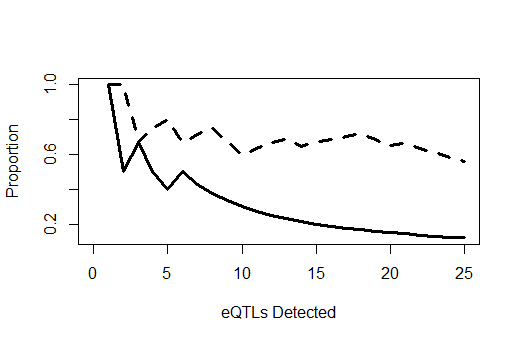}
	\end{tabular}
\end{figure}

To further examine the eQTL detection by HC-LORS and MS-LORS, we compare our findings with the \textit{cis} and \textit{trans} identification in the seeQTL database of the University of North Carolina at Chapel Hill, available at \url{http://www.bios.unc.edu/research/genomic_software/seeQTL/} \citep{seeQTL}, which report eQTLs identified from a
meta-analysis from HapMap human lymphoblastoid cell lines. We find that out of the 158 SNP-probe associations detected by HC-LORS, 23 are confirmed by the seeQTL database.  On the other hand, out of the 117 associations detected by MS-LORS, only 3 are confirmed by the seeQTL database.  
Figure \ref{fig:data_verify} presents the percentage of identified eQTLs that also appear in seeQTL database for each method. 
It is clearly shown that HC-LORS achieves higher percentage and, therefore, better consistency with the database. This is particularly obvious when comparing the overlapping percentage of the top 25 ranked eQTLs detected by each method (right panel of Figure \ref{fig:data_verify}). This result further supports the efficiency of our method in detecting true eQTLs and prioritizing them over noise.     

Table \ref{tab:data_verify} summarizes the identified SNPs that are also in seeQTL, the genes that the SNPs are associated with, and {\it cis/trans} classifications.  
It can be seen that, although both methods detect many \textit{trans}-eQTLs, almost all the overlapped eQTLs are {\it cis}. This could be due to the limited record of {\it trans}-eQTLs in the database. The proposed study has the potential to provide timely development for {\it trans} detection in eQTL mapping.   


\begin{table}[ht]
	\caption{SNP gene pairs identified by HC-LORS and LORS also found in the UNC seeQTL database.  * denotes that the association was found through HC-LORS or LORS and + denotes that the association was found in the seeQLT database} \label{tab:data_verify}
	\begin{tabular}{llll}
		\hline
		\textbf{Method} & \textbf{SNP} & \textbf{Associated Gene(s)} & \textbf{Classification}  \\ \hline
		\multirow{15}{*}{HC-LORS} &rs10888391 & $CTSS^{*+}$ & cis \\ 
		& rs11204737 & $CTSS^{*+}$ & cis \\ 
		& rs12137269 & $ST7L^{*+}$ & cis  \\ 
		& rs12568757 & $CTSK^{*+}$ & cis  \\ 
		& rs12745189 & $GSTM1^{*+} GSTM4^{*+}$ & cis  \\ 
		& rs2138686 & $OXCT2^{*+} PPIE^{*+} BMP8B^{*+}$ & cis   \\ 
		& rs2297663 & $C1orf123^{*+}$ & cis  \\ 
		& rs234098 & $FAM129A^{*+}$ & cis   \\ 
		& rs3007708 & $S100A10^{*+} THEM4^{*+}$ & cis  \\ 
		& rs4654748 & $NBPF3^{*+}$ & cis  \\ 
		& rs4660652 & $NDUFS5^{*+}$ & cis  \\ 
		& rs518365 & $IPP^{*+}$ & cis   \\ 
		& rs6684005 & $EFCAB2^{*+}$ & cis   \\ 
		& rs954679 & $ST7L^{*+}$ & cis  \\ 
		& rs2310752 & $TGFBR3^* DPYD^+$ & trans\\ \hline 
		\multirow{2}{*}{LORS} & rs2239892 & $GSTM1^{*+}$ & cis\\ 
		& rs4926440 & $SCCPDH^{*+}$ & cis \\ \hline
	\end{tabular}
\end{table}


\section{Discussion}

In this paper, we present a new ranking and screening strategy based on the higher criticism (HC) statistic for eQTL mapping.   The HC ranking and screening is effective in prioritizing and selecting both strong-sparse and weak-dense SNPs over noise SNPs. Combing HC ranking and screening with joint modeling, our multi-step procedure, HC-LORS, shows higher accuracy in detecting true eQTLs with lower computation cost compared to existing methods. 
In an analysis of eQTL data from the International HapMap Project, the result of HC-LORS shows higher consistency with the seeQTL database than existing methods. 

Due to limited sample size ($n=160$), SNPs on  chromosome 1 have been included in our analysis. The proposed method can be applied to genome-wide eQTL mapping with larger sample size.  
The emerging technological discoveries will result in more data being available for eQTL analysis and greater computational challenge.  Because of this, an effective screening method to reduce the number of SNPs will be of even greater importance.  Our method shows promise in detecting {\it cis}- and \textit{trans}-eQTLs and  dramatically reducing computational expense.

\section*{Acknowledgment}

Dr. Jeng was partially supported by National Human Genome Research Institute of the National Institute of Health under grant R03HG008642. Dr. Tzeng was partially supported by National Institutes of Health under grant P01 CA142538.

\section*{Appendix}

\subsection*{Marginal Estimate Algorithm}

We obtain the solution of (\ref{LORS_Screening_optimization}) using the screening algorithm presented in the supplemental document of \cite{yang2013LORS} based on the following Lemma from \cite{Mazumder}.  

\begin{lemma}
	\label{Mazudmer_Lemma}
	Suppose matrix \textbf{W} has rank $r$.  The solution to the optimization problem $min_{\textbf{Z}} \lVert \textbf{W} - \textbf{Z} \rVert_F^2 + \lambda \lVert \textbf{Z} \rVert_*$ is given by $\hat{\textbf{Z}} = S_\lambda (\textbf{W})$ where $S_\lambda (\textbf{W}) = \textbf{U}\textbf{D}_\lambda \textbf{V}^T$ with $\textbf{D}_\lambda = diag[(d_1 - \lambda)_+,...,(d_r - \lambda)_+]$ and $\textbf{UDV}^T$ is the SVD of \textbf{W}
\end{lemma}
	
The screening algorithm iteratively solve the follows. 
	
	\begin{enumerate}
		\item Fix $(\textbf{B}_i, \mu)$.  The optimization problem becomes $min_{\textbf{L}} || \textbf{Y} - \textbf{X}_i\textbf{B}_i - \textbf{1}\textbf{$\mu$} - \textbf{L}||^2_F +  \lambda||\textbf{L}||_*$
		\begin{itemize}
			\item By Lemma \ref{Mazudmer_Lemma} the solution is $\textbf{L} = S_{\lambda}(\textbf{Y} - \textbf{X}_i \textbf{B}_i - \textbf{1}\mu)$
			\end{itemize}
			\item Fix $\textbf{L}$.  The optimization problem becomes $argmin_{(\textbf{B}_i,\mu)} \frac{1}{2} \lVert \textbf{Y} - \textbf{L} - \textbf{X}_i\textbf{B}_i - \textbf{1}\mu \rVert_2^2$
			\begin{itemize}
				\item This problem can be solved by ordinary least squares
				\end{itemize}
				\end{enumerate}




\end{document}


	

	\newcommand{\blind}{1}

	

	\if0\blind
	{
		\title{\bf Supplemental Document for eQTL Mapping via Effective SNP
			Ranking and Screening \thanks{Address for correspondence: X. Jessie Jeng, Department of Statistics, North Carolina State University, SAS Hall, 2311 Stinson Dr., Raleigh, NC 27695-8203, USA. E-mail: xjjeng@ncsu.edu. 
			}\hspace{.2cm}}
		\author{Jacob Rhyne$^1$, Jung-Ying Tzeng$^{1,2,3,4}$, Teng Zhang$^1$, and X. Jessie Jeng$^1$ \\ 
			1. Department of Statistics, North Carolina State University\\
			2. Bioinformatics Research Center, North Carolina State University \\
			3. Department of Statistics, National Cheng-Kung University, Tainan, Taiwan. \\
			4. Institute of Epidemiology and Preventive Medicine,\\ 
			National Taiwan University, Taipei, Taiwan\\
		}
		\maketitle
	} \fi

	\if1\blind
	{
		\bigskip
		\bigskip
		\bigskip
		\begin{center}
			{\LARGE\bf Supplemental Document for eQTL Mapping via Effective SNP Ranking and Screening \\}
		\end{center}
		\medskip
	} \fi
	
	\hspace{-0.5in}
	

\begin{abstract}
This document contains additional tables for \it{eQTL Mapping via Effective SNP Ranking and Screening}
\end{abstract}




\section*{Full list of associations detected by HC-LORS}


\begin{longtable}{lll}
\hline
\textbf{SNP (Gene)} & \textbf{Probe (Gene)} & \textbf{Classification} \\ \hline
rs12745189 (Intergenic) & ILMN\_1762255 (GSTM1) & cis\\ \hline
rs12745189 (Intergenic) & ILMN\_1668134 (GSTM1) & cis\\ \hline
rs4657741 (TIPRL) & ILMN\_1779432 (TIPRL) & cis\\ \hline
rs2138686 (BMP8B) & ILMN\_1653730 (OXCT2) & cis\\ \hline
rs3007708 (Intergenic) & ILMN\_1796712 (S100A10) & cis\\ \hline
rs883694 (Intergenic) & ILMN\_1675557 (C1orf102) & cis\\ \hline
rs518365 (Intergenic) & ILMN\_1789106 (IPP) & cis\\ \hline
rs12745189 (Intergenic) & ILMN\_1753405 (GSTM4) & cis\\ \hline
rs4491070 (TNFRSF8) & ILMN\_1762255 (GSTM1) & trans\\ \hline
rs731075 (Intergenic) & ILMN\_1769839 (L1TD1) & trans\\ \hline
rs6684005 (EFCAB2) & ILMN\_1694068 (EFCAB2) & cis\\ \hline
rs12745189 (Intergenic) & ILMN\_1651800 (GSTM4) & cis\\ \hline
rs12137269 (ST7L) & ILMN\_1659926 (ST7L) & cis\\ \hline
rs2297663 (LRP8) & ILMN\_1804339 (CAMK1G) & trans\\ \hline
rs10888391 (CTSS) & ILMN\_1743032 (CTSS) & cis\\ \hline
rs3007708 (Intergenic) & ILMN\_1682120 (THEM4) & cis\\ \hline
rs234098 (FAM129A) & ILMN\_1810725 (FAM129A) & cis\\ \hline
rs12137269 (ST7L) & ILMN\_1660738 (ST7L) & cis\\ \hline
rs234098 (FAM129A) & ILMN\_1667966 (C1orf24) & cis\\ \hline
rs3007217 (Intergenic) & ILMN\_1765459 (S100A13) & trans\\ \hline
rs4654748 (NBPF3) & ILMN\_1680348 (NBPF3) & cis\\ \hline
rs4657741 (TIPRL) & ILMN\_1781457 (TIPRL) & cis\\ \hline
rs473821 (Intergenic) & ILMN\_1765459 (S100A13) & trans\\ \hline
rs11204737 (ARNT) & ILMN\_1697519 (ARNT) & cis\\ \hline
rs6704100 (Intergenic) & ILMN\_1769839 (L1TD1) & trans\\ \hline
rs6428558 (Intergenic) & ILMN\_1702231 (C1orf54) & trans\\ \hline
rs4491070 (TNFRSF8) & ILMN\_1668134 (GSTM1) & trans\\ \hline
rs2297663 (LRP8) & ILMN\_1769839 (L1TD1) & trans\\ \hline
rs10920542 (Intergenic) & ILMN\_1804339 (CAMK1G) & trans\\ \hline
rs12745189 (Intergenic) & ILMN\_1806165 (HSPA6) & trans\\ \hline
rs954679 (RHOC) & ILMN\_1659926 (ST7L) & cis\\ \hline
rs6428558 (Intergenic) & ILMN\_1726030 (GPX7) & trans\\ \hline
rs883694 (Intergenic) & ILMN\_1719835 (C1orf102) & cis\\ \hline
rs10888391 (CTSS) & ILMN\_1697519 (ARNT) & cis\\ \hline
rs12408890 (WDR8) & ILMN\_1711166 (WDR8) & cis\\ \hline
rs12563583 (Intergenic) & ILMN\_1659975 (FLJ38984) & cis\\ \hline
rs6691537 (RYR2) & ILMN\_1806165 (HSPA6) & trans\\ \hline
rs1434378 (Intergenic) & ILMN\_1782389 (LAD1) & trans\\ \hline
rs12408930 (FMN2) & ILMN\_1669790 (MNDA) & trans\\ \hline
rs274741 (Intergenic) & ILMN\_1809751 (EIF2C3) & cis\\ \hline
rs3754076 (Intergenic) & ILMN\_1784287 (TGFBR3) & trans\\ \hline
rs12033037 (Intergenic) & ILMN\_1697267 (PRKCZ) & trans\\ \hline
rs6693817 (PLEKHO1) & ILMN\_1685387 (PIGR) & trans\\ \hline
rs2376487 (Intergenic) & ILMN\_1732198 (UTS2) & trans\\ \hline
rs12568757 (CTSS) & ILMN\_1758895 (CTSK) & cis\\ \hline
rs2138686 (BMP8B) & ILMN\_1680341 (PPIE) & cis\\ \hline
rs6658304 (Intergenic) & ILMN\_1669790 (MNDA) & trans\\ \hline
rs473821 (Intergenic) & ILMN\_1669790 (MNDA) & trans\\ \hline
rs954679 (RHOC) & ILMN\_1660738 (ST7L) & cis\\ \hline
rs1073940 (DNM3) & ILMN\_1763638 (BCAR3) & trans\\ \hline
rs731075 (Intergenic) & ILMN\_1765459 (S100A13) & trans\\ \hline
rs6704100 (Intergenic) & ILMN\_1804339 (CAMK1G) & trans\\ \hline
rs883694 (Intergenic) & ILMN\_1784287 (TGFBR3) & trans\\ \hline
rs4379719 (NAV1) & ILMN\_1656186 (SLC41A1) & semi-cis\\ \hline
rs9661248 (Intergenic) & ILMN\_1739496 (PRRX1) & trans\\ \hline
rs6428558 (Intergenic) & ILMN\_1665775 (MOSC2) & trans\\ \hline
rs10793715 (Intergenic) & ILMN\_1739496 (PRRX1) & trans\\ \hline
rs12034254 (Intergenic) & ILMN\_1721580 (TBX15) & trans\\ \hline
rs11204737 (ARNT) & ILMN\_1739496 (PRRX1) & trans\\ \hline
rs3790606 (WNT2B) & ILMN\_1806165 (HSPA6) & trans\\ \hline
rs6427641 (IL6R) & ILMN\_1656415 (CDKN2C) & trans\\ \hline
rs519553 (SYT6) & ILMN\_1713952 (C1orf106) & trans\\ \hline
rs2297663 (LRP8) & ILMN\_1713053 (C1orf123) & cis\\ \hline
rs4656349 (NOS1AP) & ILMN\_1765459 (S100A13) & trans\\ \hline
rs12568757 (CTSS) & ILMN\_1765459 (S100A13) & semi-cis\\ \hline
rs2297663 (LRP8) & ILMN\_1681301 (AIM2) & trans\\ \hline
rs645383 (EIF2C1) & ILMN\_1674985 (TMEM51) & trans\\ \hline
rs645383 (EIF2C1) & ILMN\_1809751 (EIF2C3) & cis\\ \hline
rs349429 (HIVEP3) & ILMN\_1663131 (LYST) & trans\\ \hline
rs10918840 (NOS1AP) & ILMN\_1784287 (TGFBR3) & trans\\ \hline
rs10889531 (Intergenic) & ILMN\_1739496 (PRRX1) & trans\\ \hline
rs3007217 (Intergenic) & ILMN\_1734937 (IL23R) & trans\\ \hline
rs518365 (Intergenic) & ILMN\_1774938 (AKR1A1) & cis\\ \hline
rs2138686 (BMP8B) & ILMN\_1769782 (LAX1) & trans\\ \hline
rs3007708 (Intergenic) & ILMN\_1674985 (TMEM51) & trans\\ \hline
rs2138686 (BMP8B) & ILMN\_1815628 (BMP8B) & cis\\ \hline
rs12408890 (WDR8) & ILMN\_1727479 (FAM79A) & cis\\ \hline
rs4379719 (NAV1) & ILMN\_1691693 (FCRL3) & trans\\ \hline
rs1926264 (Intergenic) & ILMN\_1682428 (C1orf59) & trans\\ \hline
rs11204737 (ARNT) & ILMN\_1746801 (CGN) & semi-cis\\ \hline
rs12137269 (ST7L) & ILMN\_1685045 (CHI3L2) & semi-cis\\ \hline
rs485874 (Intergenic) & ILMN\_1703558 (FHL3) & trans\\ \hline
rs3754076 (Intergenic) & ILMN\_1697448 (TXNIP) & trans\\ \hline
rs1370722 (Intergenic) & ILMN\_1697448 (TXNIP) & trans\\ \hline
rs2404393 (Intergenic) & ILMN\_1807136 (LOC729559) & trans\\ \hline
rs4379719 (NAV1) & ILMN\_1782389 (LAD1) & semi-cis\\ \hline
rs4656349 (NOS1AP) & ILMN\_1654983 (LDLRAD2) & trans\\ \hline
rs10918840 (NOS1AP) & ILMN\_1718863 (KCNK1) & trans\\ \hline
rs4654714 (Intergenic) & ILMN\_1680393 (SNORD55) & trans\\ \hline
rs12139364 (TESK2) & ILMN\_1789106 (IPP) & semi-cis\\ \hline
rs10924734 (Intergenic) & ILMN\_1795839 (SCCPDH) & semi-cis\\ \hline
rs1370722 (Intergenic) & ILMN\_1707088 (DENND2D) & trans\\ \hline
rs3007708 (Intergenic) & ILMN\_1705078 (CHI3L2) & trans\\ \hline
rs6541199 (Intergenic) & ILMN\_1685387 (PIGR) & trans\\ \hline
rs11204737 (ARNT) & ILMN\_1743032 (CTSS) & cis\\ \hline
rs12031354 (Intergenic) & ILMN\_1784287 (TGFBR3) & trans\\ \hline
rs4656349 (NOS1AP) & ILMN\_1718863 (KCNK1) & trans\\ \hline
rs9661248 (Intergenic) & ILMN\_1748992 (RHOU) & trans\\ \hline
rs12033037 (Intergenic) & ILMN\_1784287 (TGFBR3) & trans\\ \hline
rs6667186 (Intergenic) & ILMN\_1754894 (C1orf162) & trans\\ \hline
rs6691537 (RYR2) & ILMN\_1685387 (PIGR) & trans\\ \hline
rs12033037 (Intergenic) & ILMN\_1748992 (RHOU) & trans\\ \hline
rs6667186 (Intergenic) & ILMN\_1703558 (FHL3) & trans\\ \hline
rs4660652 (AKIRIN1) & ILMN\_1776104 (NDUFS5) & cis\\ \hline
rs7525018 (DPYD) & ILMN\_1654983 (LDLRAD2) & trans\\ \hline
rs3007217 (Intergenic) & ILMN\_1671568 (ECHDC2) & trans\\ \hline
rs1317002 (SMYD3) & ILMN\_1784287 (TGFBR3) & trans\\ \hline
rs1369848 (CNIH3) & ILMN\_1653730 (OXCT2) & trans\\ \hline
rs7543224 (TBX19) & ILMN\_1806165 (HSPA6) & trans\\ \hline
rs12034383 (CR1) & ILMN\_1804339 (CAMK1G) & semi-cis\\ \hline
rs518365 (Intergenic) & ILMN\_1723684 (DARC) & trans\\ \hline
rs6691208 (Intergenic) & ILMN\_1762255 (GSTM1) & trans\\ \hline
rs2297663 (LRP8) & ILMN\_1766319 (SH2D2A) & trans\\ \hline
rs12139364 (TESK2) & ILMN\_1803005 (MMACHC) & cis\\ \hline
rs7522283 (PPM1J) & ILMN\_1769839 (L1TD1) & trans\\ \hline
rs6658304 (Intergenic) & ILMN\_1752639 (SLC25A24) & trans\\ \hline
rs1926264 (Intergenic) & ILMN\_1705078 (CHI3L2) & trans\\ \hline
rs4654748 (NBPF3) & ILMN\_1771385 (GBP4) & trans\\ \hline
rs2306459 (SLC1A7) & ILMN\_1769782 (LAX1) & trans\\ \hline
rs12408890 (WDR8) & ILMN\_1769839 (L1TD1) & trans\\ \hline
rs10920828 (Intergenic) & ILMN\_1653730 (OXCT2) & trans\\ \hline
rs10799576 (Intergenic) & ILMN\_1782389 (LAD1) & trans\\ \hline
rs12408890 (WDR8) & ILMN\_1726030 (GPX7) & trans\\ \hline
rs6684005 (EFCAB2) & ILMN\_1710974 (RGS2) & trans\\ \hline
rs7545963 (Intergenic) & ILMN\_1726030 (GPX7) & trans\\ \hline
rs12735611 (Intergenic) & ILMN\_1743455 (IL12RB2) & trans\\ \hline
rs2788557 (Intergenic) & ILMN\_1804339 (CAMK1G) & trans\\ \hline
rs2297663 (LRP8) & ILMN\_1762255 (GSTM1) & trans\\ \hline
rs4911997 (CAPZB) & ILMN\_1682428 (C1orf59) & trans\\ \hline
rs12408890 (WDR8) & ILMN\_1762255 (GSTM1) & trans\\ \hline
rs1369848 (CNIH3) & ILMN\_1754894 (C1orf162) & trans\\ \hline
rs1780321 (Intergenic) & ILMN\_1804339 (CAMK1G) & trans\\ \hline
rs6691537 (RYR2) & ILMN\_1705078 (CHI3L2) & trans\\ \hline
rs1369848 (CNIH3) & ILMN\_1743455 (IL12RB2) & trans\\ \hline
rs6428558 (Intergenic) & ILMN\_1721580 (TBX15) & trans\\ \hline
rs4911997 (CAPZB) & ILMN\_1702231 (C1orf54) & trans\\ \hline
rs12139364 (TESK2) & ILMN\_1774938 (AKR1A1) & cis\\ \hline
rs1780321 (Intergenic) & ILMN\_1739496 (PRRX1) & trans\\ \hline
rs12735611 (Intergenic) & ILMN\_1657446 (C1orf57) & trans\\ \hline
rs10494816 (Intergenic) & ILMN\_1763638 (BCAR3) & trans\\ \hline
rs4654714 (Intergenic) & ILMN\_1682402 (SNORD46) & trans\\ \hline
rs6658304 (Intergenic) & ILMN\_1762255 (GSTM1) & trans\\ \hline
rs10889531 (Intergenic) & ILMN\_1654983 (LDLRAD2) & trans\\ \hline
rs2310752 (PDE4B) & ILMN\_1784287 (TGFBR3) & trans\\ \hline
rs2297663 (LRP8) & ILMN\_1704537 (PHGDH) & trans\\ \hline
rs10918840 (NOS1AP) & ILMN\_1705078 (CHI3L2) & trans\\ \hline
rs10889531 (Intergenic) & ILMN\_1754894 (C1orf162) & trans\\ \hline
rs7525018 (DPYD) & ILMN\_1732198 (UTS2) & trans\\ \hline
rs11204737 (ARNT) & ILMN\_1669790 (MNDA) & trans\\ \hline
rs1434378 (Intergenic) & ILMN\_1726030 (GPX7) & trans\\ \hline
rs3790606 (WNT2B) & ILMN\_1714170 (SPSB1) & trans\\ \hline
rs12408890 (WDR8) & ILMN\_1765855 (LRRC8C) & trans\\ \hline
rs7525018 (DPYD) & ILMN\_1754894 (C1orf162) & trans\\ \hline
rs6684005 (EFCAB2) & ILMN\_1782439 (CNN3) & trans\\ \hline
rs6658304 (Intergenic) & ILMN\_1806015 (LOC391045) & trans\\ \hline
rs6667186 (Intergenic) & ILMN\_1781374 (TUFT1) & trans\\ \hline
rs10159067 (Intergenic) & ILMN\_1674985 (TMEM51) & trans\\ \hline
rs6682184 (Intergenic) & ILMN\_1769839 (L1TD1) & trans\\ \hline
\end{longtable}

\section*{Full list of associations detected by MS-LORS, the procedure in \cite{yang2013LORS}}

\begin{longtable}{lll} 
\hline
\textbf{SNP (Gene)} & \textbf{Probe (Gene)} & \textbf{Classification} \\ \hline
rs2239892 (GSTM1) & ILMN\_1762255 (GSTM1) & cis\\ \hline
rs10802457 (ZNF695) & ILMN\_1705078 (CHI3L2) & trans\\ \hline
rs2239892 (GSTM1) & ILMN\_1668134 (GSTM1) & cis\\ \hline
rs10802457 (ZNF695) & ILMN\_1685045 (CHI3L2) & trans\\ \hline
rs12121466 (Intergenic) & ILMN\_1769839 (L1TD1) & trans\\ \hline
rs4926440 (SCCPDH) & ILMN\_1795839 (SCCPDH) & cis\\ \hline
rs10800431 (NME7) & ILMN\_1736862 (ATP1B1) & cis\\ \hline
rs2495510 (Intergenic) & ILMN\_1769839 (L1TD1) & trans\\ \hline
rs11578440 (Intergenic) & ILMN\_1782389 (LAD1) & trans\\ \hline
rs10800431 (NME7) & ILMN\_1782389 (LAD1) & trans\\ \hline
rs41519945 (PLA2G4A) & ILMN\_1705078 (CHI3L2) & trans\\ \hline
rs1128750 (KLHDC9) & ILMN\_1691693 (FCRL3) & semi-cis\\ \hline
rs556585 (SLC35F3) & ILMN\_1663318 (RGS13) & trans\\ \hline
rs2495053 (PRDM2) & ILMN\_1769839 (L1TD1) & trans\\ \hline
rs4926440 (SCCPDH) & ILMN\_1769839 (L1TD1) & trans\\ \hline
rs683803 (Intergenic) & ILMN\_1785493 (FCER1G) & trans\\ \hline
rs16851997 (TMEM82) & ILMN\_1732198 (UTS2) & trans\\ \hline
rs12087657 (CKS1B) & ILMN\_1739496 (PRRX1) & trans\\ \hline
rs17536126 (Intergenic) & ILMN\_1769839 (L1TD1) & semi-cis\\ \hline
rs6690609 (QSOX1) & ILMN\_1663318 (RGS13) & trans\\ \hline
rs705191 (Intergenic) & ILMN\_1785493 (FCER1G) & trans\\ \hline
rs9308476 (Intergenic) & ILMN\_1704291 (LOC645317) & trans\\ \hline
rs12058761 (Intergenic) & ILMN\_1674985 (TMEM51) & trans\\ \hline
rs4522021 (DAB1) & ILMN\_1748992 (RHOU) & trans\\ \hline
rs9426844 (Intergenic) & ILMN\_1756953 (GBP6) & trans\\ \hline
rs11120919 (CAMTA1) & ILMN\_1691693 (FCRL3) & trans\\ \hline
rs1316855 (Intergenic) & ILMN\_1739496 (PRRX1) & trans\\ \hline
rs6577582 (NOL9) & ILMN\_1782439 (CNN3) & trans\\ \hline
rs6670639 (FMO1) & ILMN\_1726030 (GPX7) & trans\\ \hline
rs11801847 (Intergenic) & ILMN\_1782389 (LAD1) & trans\\ \hline
rs17460402 (Intergenic) & ILMN\_1769839 (L1TD1) & trans\\ \hline
rs4926440 (SCCPDH) & ILMN\_1782439 (CNN3) & trans\\ \hline
rs11806030 (Intergenic) & ILMN\_1663318 (RGS13) & semi-cis\\ \hline
rs2000321 (ATP1B1) & ILMN\_1736862 (ATP1B1) & cis\\ \hline
rs2454290 (HSPG2) & ILMN\_1663318 (RGS13) & trans\\ \hline
rs4233478 (Intergenic) & ILMN\_1739496 (PRRX1) & trans\\ \hline
rs12078747 (INADL) & ILMN\_1694068 (EFCAB2) & trans\\ \hline
rs2485653 (KLF17) & ILMN\_1738517 (FCRL4) & trans\\ \hline
rs6663219 (Intergenic) & ILMN\_1806165 (HSPA6) & trans\\ \hline
rs1868302 (Intergenic) & ILMN\_1704291 (LOC645317) & trans\\ \hline
rs12029613 (Intergenic) & ILMN\_1665775 (MOSC2) & trans\\ \hline
rs12752801 (Intergenic) & ILMN\_1663318 (RGS13) & trans\\ \hline
rs2239892 (GSTM1) & ILMN\_1663131 (LYST) & trans\\ \hline
rs4926440 (SCCPDH) & ILMN\_1685387 (PIGR) & trans\\ \hline
rs2498967 (INADL) & ILMN\_1741944 (SNORD38B) & trans\\ \hline
rs4657165 (NOS1AP) & ILMN\_1807529 (PADI4) & trans\\ \hline
rs2281182 (AOF2) & ILMN\_1739496 (PRRX1) & trans\\ \hline
rs4926440 (SCCPDH) & ILMN\_1686968 (FLJ25476) & trans\\ \hline
rs2498967 (INADL) & ILMN\_1797905 (HIST2H4A) & trans\\ \hline
rs2239892 (GSTM1) & ILMN\_1697267 (PRKCZ) & trans\\ \hline
rs4926440 (SCCPDH) & ILMN\_1736862 (ATP1B1) & trans\\ \hline
rs4522021 (DAB1) & ILMN\_1691693 (FCRL3) & trans\\ \hline
rs4926440 (SCCPDH) & ILMN\_1795461 (LOC729853) & trans\\ \hline
rs4926440 (SCCPDH) & ILMN\_1694539 (MAP3K6) & trans\\ \hline
rs4926440 (SCCPDH) & ILMN\_1789436 (FLJ20054) & trans\\ \hline
rs2239892 (GSTM1) & ILMN\_1784287 (TGFBR3) & trans\\ \hline
rs4926440 (SCCPDH) & ILMN\_1702231 (C1orf54) & trans\\ \hline
rs1128750 (KLHDC9) & ILMN\_1797428 (FCRL3) & semi-cis\\ \hline
rs3766667 (CEP170) & ILMN\_1691693 (FCRL3) & trans\\ \hline
rs2239892 (GSTM1) & ILMN\_1718863 (KCNK1) & trans\\ \hline
rs4926440 (SCCPDH) & ILMN\_1765855 (LRRC8C) & trans\\ \hline
rs4926440 (SCCPDH) & ILMN\_1682081 (IBRDC3) & trans\\ \hline
rs11803446 (Intergenic) & ILMN\_1684278 (SNORD38A) & trans\\ \hline
rs9428859 (RGS7) & ILMN\_1682428 (C1orf59) & trans\\ \hline
rs2495053 (PRDM2) & ILMN\_1656186 (SLC41A1) & trans\\ \hline
rs2239892 (GSTM1) & ILMN\_1754894 (C1orf162) & semi-cis\\ \hline
rs1001451 (Intergenic) & ILMN\_1796712 (S100A10) & trans\\ \hline
rs4926440 (SCCPDH) & ILMN\_1731692 (LOC391157) & trans\\ \hline
rs2239892 (GSTM1) & ILMN\_1671337 (SLC2A5) & trans\\ \hline
rs4926440 (SCCPDH) & ILMN\_1668134 (GSTM1) & trans\\ \hline
rs11120919 (CAMTA1) & ILMN\_1682428 (C1orf59) & trans\\ \hline
rs4926440 (SCCPDH) & ILMN\_1782070 (NPL) & trans\\ \hline
rs4657165 (NOS1AP) & ILMN\_1785493 (FCER1G) & semi-cis\\ \hline
rs4522021 (DAB1) & ILMN\_1665775 (MOSC2) & trans\\ \hline
rs4926440 (SCCPDH) & ILMN\_1749915 (C1orf63) & trans\\ \hline
rs4926440 (SCCPDH) & ILMN\_1789106 (IPP) & trans\\ \hline
rs4926440 (SCCPDH) & ILMN\_1692698 (VASH2) & trans\\ \hline
rs4926440 (SCCPDH) & ILMN\_1695583 (LOC644094) & trans\\ \hline
rs4926440 (SCCPDH) & ILMN\_1660027 (FCGR2B) & trans\\ \hline
rs1868302 (Intergenic) & ILMN\_1748992 (RHOU) & trans\\ \hline
rs4926440 (SCCPDH) & ILMN\_1746801 (CGN) & trans\\ \hline
rs11120919 (CAMTA1) & ILMN\_1702231 (C1orf54) & trans\\ \hline
rs4926440 (SCCPDH) & ILMN\_1806607 (SFN) & trans\\ \hline
rs4926440 (SCCPDH) & ILMN\_1807529 (PADI4) & trans\\ \hline
rs4926440 (SCCPDH) & ILMN\_1691611 (LOC645436) & trans\\ \hline
rs4926440 (SCCPDH) & ILMN\_1672295 (ZC3H12A) & trans\\ \hline
rs4926440 (SCCPDH) & ILMN\_1723684 (DARC) & trans\\ \hline
rs11805614 (Intergenic) & ILMN\_1682428 (C1orf59) & trans\\ \hline
rs4926440 (SCCPDH) & ILMN\_1687998 (LPGAT1) & trans\\ \hline
rs4926440 (SCCPDH) & ILMN\_1707088 (DENND2D) & trans\\ \hline
rs4926440 (SCCPDH) & ILMN\_1815700 (WNT3A) & trans\\ \hline
rs10800431 (NME7) & ILMN\_1734937 (IL23R) & trans\\ \hline
rs4926440 (SCCPDH) & ILMN\_1711031 (RALGPS2) & trans\\ \hline
rs4926440 (SCCPDH) & ILMN\_1798284 (MFSD2) & trans\\ \hline
rs2239892 (GSTM1) & ILMN\_1782389 (LAD1) & trans\\ \hline
rs4926440 (SCCPDH) & ILMN\_1711166 (WDR8) & trans\\ \hline
rs4926440 (SCCPDH) & ILMN\_1806165 (HSPA6) & trans\\ \hline
rs11584748 (Intergenic) & ILMN\_1653730 (OXCT2) & trans\\ \hline
rs2239892 (GSTM1) & ILMN\_1789436 (FLJ20054) & trans\\ \hline
rs2239892 (GSTM1) & ILMN\_1686968 (FLJ25476) & trans\\ \hline
rs4926440 (SCCPDH) & ILMN\_1746175 (TNFSF4) & trans\\ \hline
rs12132905 (Intergenic) & ILMN\_1782389 (LAD1) & trans\\ \hline
rs4926440 (SCCPDH) & ILMN\_1783226 (SSR2) & trans\\ \hline
rs4926440 (SCCPDH) & ILMN\_1763638 (BCAR3) & trans\\ \hline
rs4926440 (SCCPDH) & ILMN\_1762255 (GSTM1) & trans\\ \hline
rs4926440 (SCCPDH) & ILMN\_1776519 (RAP1GAP) & trans\\ \hline
rs4926440 (SCCPDH) & ILMN\_1658071 (ATP1B1) & trans\\ \hline
rs4926440 (SCCPDH) & ILMN\_1734937 (IL23R) & trans\\ \hline
rs17124724 (Intergenic) & ILMN\_1804339 (CAMK1G) & trans\\ \hline
rs1294248 (KIAA1804) & ILMN\_1796712 (S100A10) & trans\\ \hline
rs4926440 (SCCPDH) & ILMN\_1680579 (ATP2B4) & trans\\ \hline
rs4926440 (SCCPDH) & ILMN\_1653504 (EDG1) & trans\\ \hline
rs4926440 (SCCPDH) & ILMN\_1749011 (NECAP2) & trans\\ \hline
rs1332963 (Intergenic) & ILMN\_1702534 (CD244) & trans\\ \hline
rs12736336 (Intergenic) & ILMN\_1752639 (SLC25A24) & trans\\ \hline
rs4658753 (KIF26B) & ILMN\_1691693 (FCRL3) & trans\\ \hline
rs4926440 (SCCPDH) & ILMN\_1803818 (NMNAT2) & trans\\ \hline
\end{longtable}